\documentclass[letterpaper,10pt]{article}
\usepackage{osameet2}

\usepackage{amsmath,amssymb}
\usepackage{hyperref}
\hypersetup{colorlinks=true,citecolor=blue,urlcolor=blue}

\begin{document}

\title{Partially Ordered Statistics Demapping for Multi-Dimensional Modulation Formats}

\author{Djalal F. Bendimerad, Huijian Zhang, Ingmar Land, and Hartmut Hafermann}
\address{Mathematical and Algorithmic Sciences Lab, Paris Research Center, Huawei Technologies France SASU}
\email{djalal.falih.bendimerad@huawei.com}

\begin{abstract}
We propose a very low-complexity and high-performance algorithm for soft-demapping of multi-dimensional modulation formats. We assess its performance over the linear channel for four 8D formats, generated using binary arithmetics. This solution outperforms current algorithms in terms of complexity without loss in performances.
\end{abstract}


\section{Introduction}
Recent advances on modulation formats for optical communications showed that a Multi Dimensional (MD) design increases the performance in both linear and nonlinear channels \cite{Reimer2016, Bendimerad2018}, compared to conventional formats such as Polarization-Division-Multiplexing Quadrature-Phase-Shift-Keying (PDM-QPSK). However, the implementation complexity of the soft-demapper increases significantly, and this limits the use of these formats in optical transmission systems that employ Forward Error Correction (FEC) codes under soft-decision decoding. To address this issue, we proposed in \cite{Bendimerad2018b} to use the Min-Sum (MS) algorithm as an ultra-low complexity soft-demapper for MD formats that are generated using binary arithmetics, referred to as Boolean Equations (BEs).
This algorithm operates on the Tanner graph, and can be used for formats that are generated using quasi-linear BEs with the property that each bit occurs only once, and therefore, can be expressed as a function of the other bits. In other words, observations at the demapper are not correlated, and extrinsic information can be easily identified\cite{Bendimerad2018b}. For modulation formats that are generated using nonlinear BEs\cite{Bendimerad2018}, where all observations at the demapper are correlated, it is not possible to extract the extrinsic information, and therefore, the MS algorithm cannot be used.\\
In this paper, we propose to use the Ordered Statistics Decoding (OSD) algorithm, first introduced in \cite{Fossorier1995}, for the demapping. This algorithm offers an ultra-low complexity solution with high performances, and can demap any MD format, as long as the latter is generated using Boolean Equations. Four nonlinearity-tolerant modulation formats, based on Set-Partitioned PDM-QPSK, are studied over the Additive White Gaussian Noise (AWGN) channel, and the algorithm performance is assessed using the post-FEC Bit Error Rate (BER).

\section{Principle of Operations of the pOSD Soft-Demapper}
Compared to standard formats that have independent dimensions like PDM-QPSK for example, and for which the Log-Likelihood-Ratios (LLRs) at the soft-demapper is obtained simply by measuring the observation (1D-demapping), MD formats must be demapped in all dimensions jointly. As a consequence, using observations at the demapper for an MD format without considering the BEs results on suboptimal demapping; the loss in performance can be significant, depending on the modulation format. To avoid this loss, one solution is to use the so-called MaxLogMap (MLM) algorithm to compute LLRs. However, the received observation needs to be compared to all possible MD symbols of the modulation format.
This approach is very complex, and increases significantly the cost and the power consumption of the hardware,  which makes the implementation unfeasible.\\
Instead, we propose here to use the OSD algorithm \cite{Fossorier1995}, which we adapt for the demapping. The difference here is that we order only information bits based on the absolute value of observations. We therefore refer to it as Partially Ordered Statistics Demapper (pOSD). Besides, the same generator matrix used in the mapper is used in the algorithm, which makes it simpler than the OSD where the Gaussian elimination is needed to get the new generator matrix. Basically, we choose to correct $p$ Least Reliable Positions (LRPs) out of a total of $m$ information bits. For the remaining bits positions, observations are considered as LLRs. Similarly to the Chase algorithm\cite{Chase1972}, we generate codewords using the hard decision on the observations, all possible bit combinations of the LRPs and the BEs. We then associate a metric to each candidate codeword and compute the LLR value for each LRP based on this metric.

\section{Performance of the pOSD}
\subsection{Study description}
The four modulation formats that are studied here are based on set-partitioning PDM-QPSK in 8D, i.e., sets of 4 QPSK symbols each Gray-labeled. The first and second ones are Polarization-Balanced 4 Bits in 8D (PB-4B8D, $m=4$) and PB-5B8D ($m=5$) \cite{Bendimerad2018b}, and use linear BEs in the mapping process. The third one is referred to as PB-6B8D ($m=6$). This format is equivalent to PB-QPSK\cite{Reimer2016}; we use a different name to highlight the spectral efficiency. We propose to use BEs to generate this format. Each 8D symbol is labeled by an eight-bit vector $b_1...b_8$. The first six bits are taken from the sequence of information, the last two are parity bits and computed as:
\begin{equation}
\begin{split}
b_7 = \overline{b_2}\oplus b_3 \oplus b_5\oplus \left( b_1 \oplus b_2\right) \cdot \left( b_3 \oplus b_4 \oplus b_5 \oplus b_6\right) \oplus \left( b_3 \oplus b_4 \right) \cdot \left( b_5 \oplus b_6\right)\\
b_8 = \overline{b_1} \oplus b_4 \oplus b_6 \oplus \left( b_1 \oplus b_2 \right) \cdot \left( b_3\oplus b_4\oplus b_5\oplus b_6\right) \oplus \left( b_3 \oplus b_4 \right) \cdot \left( b_5 \oplus b_6 \right)
\end{split}
\end{equation}
where [$\oplus$] and [$\cdot$] denote binary additions and multiplications, respectively, and $\overline{b}$ denotes the logical negation of $b$. By doing so, we fix its labeling, recalling that, to the best of our knowledge, no such labeling of the PB-QPSK can be found in the literature. For the complete approach on how to generate the format, please refer to \cite{Bendimerad2018}. The fourth modulation format considered is the Polarization Alternating -7B8D ($m=7$): seven bits are taken from the binary information sequence, and one parity bit is computed from a BE, as reported in \cite{Bendimerad2018}.
In order to assess the performance of the proposed soft-demapper for each modulation format, we simulate each format over the AWGN channel. The BER is calculated using a Monte Carlo loop, at the output of a Low Density Parity Check (LDPC) code decoder. We use an LDPC code of length 18K and an overhead of 20\%. As a decoder, we use a MS algorithm with 20 iterations.

\subsection{Linear Channel Performance}
\begin{figure}[htbp]
	\centering
	\includegraphics[width=17cm]{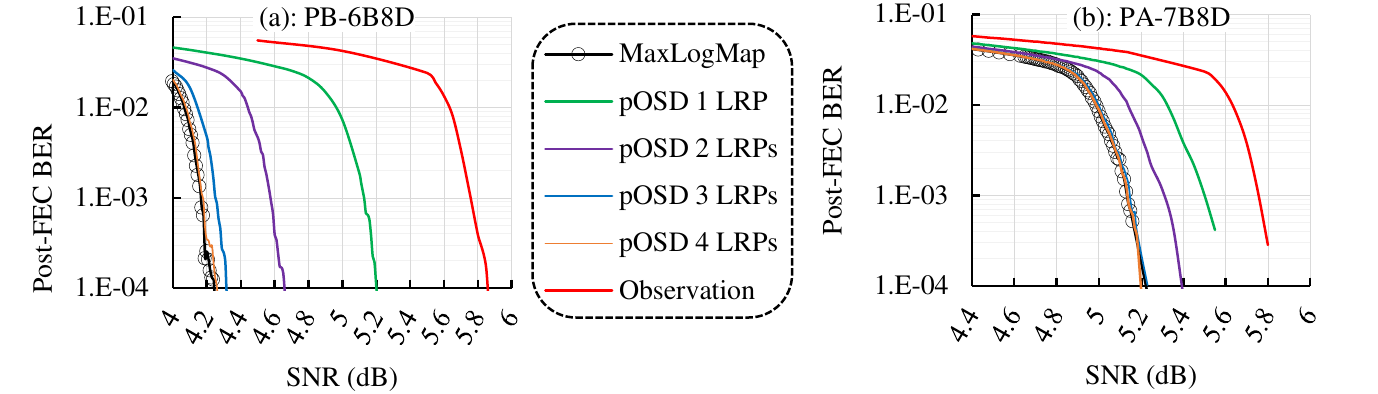}
	\caption{\label{fig:linCh}AWGN channel performance of (a) PB-6B8D and (b) PA-7B8D soft-demapping.}
\end{figure}

Figure \ref{fig:linCh} shows the linear channel performance of PB-6B8D and PA-7B8D for several soft-demapping techniques. For simplicity, we compare the curves at a BER of $3\cdot 10^{-4}$, assuming the same relative differences for lower BER values. The curve with black circles represents the optimal post-FEC BER (lower bound), which results from using the MLM algorithm as soft-demapper. The red curve represents the suboptimal post-FEC BER (upper bound), which results from considering observations as LLRs (1D-demapper).
The difference between the upper and lower bounds highlights the loss in terms of SNR. As such, losses of 1.63dB (Fig.\ref{fig:linCh}.a) and 0.63dB (Fig.\ref{fig:linCh}.b) can be seen for PB-6B8D and PA-7B8D, respectively. The two formats experience different losses because for PB-6B8D, 6 information bits must be processed using 2 BEs, while for PA-7B8D, 6 information bits are processed from only one BE. In other words, using observations as LLRs is closer to the optimal performance for PA-7B8D than for PB-6B8D. Notice that for PA-7B8D, bit 7 does not appear in the equation\cite{Bendimerad2018}, so, the observation relative to bit 7 is already an optimal LLR, as for PDM-QPSK.
It can be seen in Fig.\ref{fig:linCh} that the pOSD algorithm offers a series of alternatives to trade performance against complexity, and an optimal soft-demapping performance is achieved by processing $p=4$ LRPs for PB-6B8D (Fig.\ref{fig:linCh}.a), and $p=3$ LRPs for PA-7B8D (Fig.\ref{fig:linCh}.b). Decreasing the number of processed LRPs results in lower performances as well as lower complexity. Notice that results of PB-4B8D and PB-5B8D are not shown for lack of space. We report $k=4$ for both formats to ensure optimal performance. Finally, these relative results were validated using the achievable rate metric\cite{Bocherer2017}, in order to ensure that post-FEC BER results do not depend on a specific code.

\subsection{Complexity assessment}
We consider only the optimal version of the pOSD algorithm, namely $k=3$ LRPs for PA-7B8D and $k=4$ LRPs for PB-4B8D, PB-5B8D and PB-6B8D. The complexity assessment further depends on the sorting algorithms; we consider the merge sort algorithm\cite{Knuth1998}. The complexity of this algorithm depends on a random process, as observations at the demapper input might be already partially or completely sorted. We distinguish therefore the best and worst case scenarios for comparisons.
\begin{table}[htb]
	\centering \caption{Complexity assessment of the pOSD soft-demapper}
	\begin{tabular}{cccc|ccc|ccc|ccc}
		\hline
		\hline
		& \multicolumn{3}{c|}{PB-4B8D} & \multicolumn{3}{c|}{PB-5B8D} & \multicolumn{3}{c|}{PB-6B8D} & \multicolumn{3}{c}{PA-7B8D}  \\
		\cline{2-13}
		& MLM & MS & pOSD & MLM & MS & pOSD & MLM & MS & pOSD & MLM & MS & pOSD \\
		\cline{1-13}
		Logical op. & 260 & 210 & 304 & 654 & 71 & 304 & 1542 & $\times$ & 528 & 3078 & $\times$ & 160 \\
		Additions & 452 & 40 & 8 & 1125 & 21 & 8 & 2694 & $\times$ & 9 & 5382 & $\times$ & 4 \\
		Comparisons & 56 & 56 & 59-62 & 150 & 18 & 60-66 & 372 & $\times$ & 63-67 & 756 & $\times$ & 28-32 \\
		\# LUTs & 4 & - & - & 5 & - & - & 6 & $\times$ & - & 7 & $\times$ & - \\
		LUT size & 2$\times$8 & - & - & 2$\times$16 & - & - & 2$\times$32 & $\times$ & - & 2$\times$64 & $\times$ & - \\
		\hline
		\hline
	\end{tabular}
\label{table:comp}
\end{table}
Table \ref{table:comp} shows the type and number of operations needed for the MLM, the MS\cite{Bendimerad2018b} and the pOSD algorithms. The order of operations matters as well, as the cost of logical operations is lower than real additions, which in turn have a lower cost than real comparisons. We recall that the MS can only be applied to PB-4B8D and PB-5B8D because the other two formats are generated using nonlinear BEs. It can be observed that the pOSD outperforms the MLM, as the number of all operations is lowered by several orders of magnitude compared to the MLM. We also observe that the pOSD has similar complexity compared to the MS, reporting a slightly better performance of the pOSD for PB-4B8D and PB-5B8D as MS exhibits some losses compared to the MLM\cite{Bendimerad2018b}. Furthermore and interestingly, the pOSD soft-demapper is less complex for PA-7B8D than for all the others, although the others have a lower spectral efficiency. This is understandable by the fact that during the bit correction process, only one BE is used for the first format while several are used for the others. Finally, it is important to note that the pOSD algorithm does not require any use of Look-Up Tables (LUTs), which introduce a high complexity implementation \cite{Bendimerad2018b}.

\section{Conclusion}
In this paper, we proposed the partially ordered statistics demapping algorithm as an ultra-low complexity and low-cost soft-demapper for MD formats which are generated using Boolean equations. We showed that this algorithm does not have any performance loss compared to the MLM soft-demapper, while it outperforms the latter by reducing drastically the implementation complexity. As such, the hardware and power consumptions are significantly decreased, removing the necessity of LUTs. We also presented the Boolean equations that are used to generate the PB-6B8D, and are necessary to allow the use of the pOSD soft-demapper solution. This algorithm makes feasible the implementation of soft-demappers for formats such as PB-6B8D and PA-7B8D. This solution is appealing to be used as a unified soft-demapper for the whole MD format series. It is finally important to note that this algorithm can be applied to any MD format that uses Boolean equations in the bit-to-symbol mapping process, even for nonlinear equations.

\end{document}